\documentclass{fizik}%
\usepackage{multicol,epsfig}

\vol{27} \pyear{2003} \received{15.11.2002}
\author[BAYSAL, YILMAZ]{
\textbf{ H\"{u}sn\"{u} BAYSAL}\\
\textit{Department of Mathematics, Art and Science Faculty,}\\
\textit{\d{C}anakkale Onsekiz Mart University, 17100 \d{C}anakkale-TURKEY}\\
\textbf {\.{I}hsan YILMAZ}\\
\textit{Department of Physics, Art and Science Faculty,}\\
\textit{\d{C}anakkale Onsekiz Mart University, 17100
\d{C}anakkale-TURKEY}
}

\title{Timelike and Spacelike Ricci Collineation
Vectors in String Fluid}

\begin{document}
\maketitle

\begin{abstract}
We study the consequences of the existence of timelike and
spacelike Ricci collineation vectors (RCVs) for string fluid in
the context of general relativity. Necessary and sufficient
conditions are derived for a space-time with string fluid to admit
a timelike RCV, parallel to $u^a$, and a spacelike RCV, parallel
to $n^a$. In these cases, some results obtained are discussed.

\end{abstract}

\section{Introduction}

The introduction of a symmetry (i.e. collineation) is
most conveniently studied if the Lie derivative of the field
equations is taken with respect to the symmetry vector. More
specially, the Lie derivative of the Ricci and independently the
energy momentum tensor can be computed by the symmetry. So, the
field equations can be obtained as Lie derivatives along the
symmetry vector of the dynamical variables.

Previous works on RCVs have been undertaken by several authors.
Oliver and Davis, who gave necessary and sufficient conditions for
a matter space-time to admit an RCV, $\eta^{a} =\eta u^{a}$, with
$u^{a}=u^{a}_{D}$ where $u^{a}_{D}$ is the dynamic four-velocity
\cite{Oliver1,Oliver2}. Tsamparlis and Mason have considered Ricci
collineation vectors (RCVs) in fluid space-times (perfect,
imperfect and anisotropic) \cite{TsamMas}. Duggal have also
considered timelike Ricci inheritance vector in perfect fluid
space-times \cite{Duggal}. Carot \textit{et al.} have discussed
space times with conformal Killing vectors \cite{Carot}.

The study of string fluid has aroused considerable interest as
they are believed to give rise to density perturbations leading to
the formation of galaxies \cite{Zeldovich}. The existence of a
large scale network of strings in the early universe is not
contradiction with present day observations of the universe
\cite{Kibble}. Also, the present of strings in the early universe
can be explained using grand unified theories (GUTs)
\cite{Zeldovich,Kibble}. Thus, it is interesting to study the
symmetry features of string fluid.

Recently, work on symmetries of the string has been taken Yavuz
and Y{\i}lmaz, and Y{\i}lmaz \textit{et al.} who have considered
inheriting conformal and special conformal Killing vectors, and
also curvature inheritance symmetry in the string cosmology
(string cloud and string fluids), respectively
\cite{Yavuz,YilmazTar}. Y{\i}lmaz has also considered timelike and
spacelike Ricci collineations in the string cloud \cite{Yilmaz}.
Baysal \textit{et al.} have studied conformal collineation in the
string cosmology \cite{Baysal}.

The theory of spacelike congruences in general relativity was
first formulated by Greenberg, who also considered applications to
the vortex congruence in a rotational fluid \cite{Greenberg}. The
theory has been developed and further applications have been
considered by Mason and Tsamparlis, who also considered spacelike
conformal Killing vectors and spacelike congruences \cite{Mason}.

A space-time admits a Ricci collineation vector (RCV) $\xi^{a}$ if
\begin{equation}\label{eq1}
\pounds_{\xi} R_{ab}=0,
\end{equation}
where $R_{ab}$ is the Ricci tensor and  $\pounds_{\xi}$ denotes
Lie derivative along $\xi^{a}$. A conservation law, valid for any
RCV, was established by Collinson \cite{Collinson}. If $\xi^a$ is
an RCV, then it can be verified that
\begin{equation}\label{eq2}
(R^{ab}\xi_{b})_{;a}=0,
\end{equation}
and if Einstein's field equations

\begin{equation}\label{eq3}
R^{ab}=T^{ab}-\frac{1}{2} T g^{ab},
\end{equation}
are satisfied, then

\begin{equation}\label{eq4}
\left[(T^{ab}-\frac{1}{2} T g^{ab} )\xi_{b}\right]_{;a}=0.
\end{equation}
Equation ~(\ref{eq4}) plays an important role  as one of necessary
and sufficient conditions for a space-time to admit an RCV,
$\xi^a$.

It is important to establish a connection between timelike or spacelike RCVs
and material curves in the string fluid.

In this paper, we will investigate the properties of RCVs,
$\eta^{a} =\eta u^{a}$, parallel to the string fluid unit
four-velocity vector $u^{a}$:

\[u_{a} u^{a} =-1,\quad \eta=(-\eta_{a}\eta^{a})^{1/2}>0,\]
and spacelike RCVs, $\xi^{a} =\xi n^{a}$, orthogonal to $u^{a}$:
\[n_{a} n^{a} =+1,\quad n_a u^a=0,\quad\ \xi=(\xi_{a}\xi^{a})^{1/2}>0.\]
We will express the necessary and sufficient conditions for string fluid
space-time to admit a timelike RCV parallel to $u^{a}$ and a spacelike RCV
parallel to $n^{a}$ in terms of the kinematic quantities of the timelike
congruence of world-lines generated by $u^{a}$ and the expansion, shear, and
rotation of the spacelike congruence generated by $n^{a}$, respectively.

The energy-momentum tensor for a string fluid can be written as
\cite{Yavuz,Letelier}

\begin{equation}\label{eq5}
T_{ab}=\rho_{s}\left( u_{a} u_{b}-n_{a} n_{b}\right)+q P_{ab},
\end{equation}
where $\rho_{s}$ is string density and $q$ is "string tension" and also
"pressure".

The unit timelike vector $u^{a}$ describes the fluid four-velocity
and the unit spacelike vector  $n^{a}$ represents a direction of
anisotropy, i.e., the string's directions. Also, note that
\begin{equation}\label{eq6}
  u_{a}u^{a}=-n_{a} n^{a}=-1\quad \textrm{and} \quad u^{a}n_{a}=0.
\end{equation}

The paper may be outlined as follows. In section 2, necessary and sufficient
conditions for string fluid space-time to admit a timelike RCV parallel to
$u^a$ are derived. In section 3, the basic aspects of the theory of spacelike
congruences required in this paper are reviewed briefly. Also, necessary and
sufficient conditions for a string fluid space-time to admit a spacelike RCV
parallel to $n^a$ are given. Finally, concluding remarks are made in section 4.

\section{Timelike Ricci Collineation Vectors}

If Einstein's field equation ~(\ref{eq3}) are satisfied,
then string fluid with energy-momentum tensor ~(\ref{eq5}) admits
an RCV, $\eta^{a} =\eta u^{a}$, if and only if
\begin{equation}\label{eq7}
h^{c}_{a}h^{d}_{b}\dot\gamma_{cd}= -\frac{2}{3}\left[(2\rho_{s}-q)\sigma_{ab}
-\gamma^{cd}\sigma_{cd}h_{ab}+\theta\gamma_{ab}\right]
-2\sigma_{c(a}\gamma^{c}_{b)} -2\omega_{c(a}\gamma^{c}_{b)},
\end{equation}
\begin{equation}\label{eq8}
q\left[\dot u_{a}-(\ln\eta)_{,a}- \theta u_{a}\right]=0,
\end{equation}
\begin{equation}\label{eq9}
\left(\eta q u^{a}\right)_{;a}=0,
\end{equation}
where $\theta$ is the rate-of expansion, $\sigma_{ab}$ is the
rate-of-shear tensor, $\omega_{ab}$ is the vorticity tensor of the
timelike congruence generated by $u^{a}$,
$h_{ab}=g_{ab}+u_{a}u_{b}$ and
$\gamma_{ab}=(\rho_{s}+q)\left(\frac{1}{3}h_{ab}-n_{a}
n_{b}\right)$.

{\bf Proof:} From the definition of the Lie derivative it follows
that
\begin{equation}\label{eq10}
\pounds_{\eta u}R_{ab}=\eta\left\{\dot
R_{ab}+2u^{c}R_{c(a}(\ln\eta)_{,b)}+2R_{c(a}u^{c}_{;b)}\right\}
\end{equation}
which, using Einstein's field equation ~(\ref{eq3}) for string
fluid, may be rewritten as
\begin{eqnarray}\label{eq11}
 \pounds_{\eta u}R_{ab}&=\eta\Bigg{\{}\dot q u_{a}u_{b}+
\frac{1}{3}(2\dot\rho_{s}-\dot q)h_{ab}+\frac{4}{3}
(\rho_{s}+q)\dot
u_{(a}u_{b)}+\dot\gamma_{ab}\nonumber\\
&-2qu_{(a}(\ln\eta)_{,b)}+\frac{2}{3}(2\rho_{s}-q)u_{(a;b)}
+2\gamma_{t(a}u^{t}_{;b)}\Bigg{\}}.
\end{eqnarray}

Suppose first that $\eta u^a$  is an RCV. Then  ~(\ref{eq1}) holds
and the right-hand side of ~(\ref{eq11}) vanishes. By contracting
~(\ref{eq11}) in turn with $u^a u^b, u^a h^b_{c}, h^{ab}$, and
$h^{a}_{c}h^{b}_{d}-\frac{1}{3}h^{ab}h_{cd}$ and by using the
expansion
\begin{equation}\label{eq12}
u_{a;b}=\sigma_{ab}+\omega_{ab}+\frac{1}{3}\theta h_{ab}-\dot
u_{a}u_{b},
\end{equation}
we obtain, respectively,
\begin{eqnarray}
\dot q+2q(\ln\eta)^{.}=0,\label{eq13}\\
q\left[\dot u_{a}-(\ln\eta)_{,a}-(\ln\eta)^{.}
u_{a}\right] =0,\label{eq14}\\
2\dot\rho_{s}-\dot
q+\frac{2}{3}(2\rho_{s}-q)\theta+2\gamma^{ab}\sigma_{ab}=0,\label{eq15}
\end{eqnarray}
and equation ~(\ref{eq7}).

The energy momentum conservation equation will also be required.
For string fluid, the momentum conservation equation, which
follows from Einstein's field equations, is
\begin{equation}\label{eq16}
\dot\rho_{s}=-\frac{2}{3}(\rho_{s}+q)\theta-\gamma^{ab}\sigma_{ab}.
\end{equation}
\begin{description}
\item [(i)] Condition ~(\ref{eq7}) was derived directly in the
decomposition of ~(\ref{eq11}).

\item [(ii)] In order to determine condition ~(\ref{eq8}), we
first obtain an expression for $q(\ln\eta)^{.}$  by eliminating
$\dot\rho_{s}$ and $\dot q$ from ~(\ref{eq13}). Substituting from
~(\ref{eq16}) for $\dot\rho_{s}$ into ~(\ref{eq15}) gives
\begin{equation}\label{eq17}
\dot q = -2 q \theta,
\end{equation}
and using ~(\ref{eq16}) for $\dot\rho_{s}$ and ~(\ref{eq17})  for
$\dot q$ , equation ~(\ref{eq13}) becomes
\begin{equation}\label{eq18}
q(\ln\eta)^{.} =q\theta.
\end{equation}
Condition ~(\ref{eq8}) is derived immediately from ~(\ref{eq14})
and ~(\ref{eq18}). \item [(iii)] In order to derive condition
~(\ref{eq9}), we observe that ~(\ref{eq13}) may be written as
\begin{equation}\label{eq19}
\dot q+q(\ln\eta)^{.}+q(\ln\eta)^{.}=0.
\end{equation}
If ~(\ref{eq18}) is used to replace one of the terms
$q(\ln\eta)^{.}$ in ~(\ref{eq19}), then ~(\ref{eq19}) becomes
\begin{equation}\label{eq20}
q_{,a}\eta u^{a}+q(\eta_{,a}u^{a}+\eta u^{a}_{;a})=0,
\end{equation}
from which ~(\ref{eq9}) follows directly. Conditions
~(\ref{eq7})-~(\ref{eq9}) are therefore necessary conditions if
$\eta u^{a}$ is an RCV.
\end{description}

Conversely, suppose that conditions ~(\ref{eq7})-~(\ref{eq9}) are
satisfied. Then if ~(\ref{eq7}) for $\dot\gamma_{ab}$ is
substituted into ~(\ref{eq11}), and ~(\ref{eq12}) is used to
expand $u_{a;b}$ and $u_{t;b}$, ~(\ref{eq11}) becomes
\begin{equation}\label{eq21}
\pounds_{\eta u}R_{ab}=\eta\Bigg{\{}\left(\dot q+
2q\theta\right)u_{a}u_{b} +\frac{1}{3}\left[2\dot\rho_{s}-\dot q+
\frac{2}{3}(2\rho_{s}-q)\theta+
2\gamma^{cd}\sigma_{cd}\right]h_{ab}\Bigg{\}}.
\end{equation}
Now, $\dot\rho_{s}$ is given by the energy conservation equation
~(\ref{eq16}). In order to obtain an expression for $\dot q$, we
first observe that ~(\ref{eq9}) can be expanded as
\begin{equation}\label{eq22}
\dot q+ q\theta+q(\ln\eta)^{.}=0.
\end{equation}
But, by contracting  ~(\ref{eq8}) with $u^{a}$, ~(\ref{eq18}) is
again obtained and by eliminating $q(\ln\eta)^{.}$ from
~(\ref{eq22}), equation ~(\ref{eq17}) for $\dot q$ is again
derived.

By using ~(\ref{eq16}) for $\dot\rho_{s}$ and ~(\ref{eq17}) for
$\dot q$ it is easily verified that the coefficients of
$u_{a}u_{b}$ and $h_{ab}$ in ~(\ref{eq21}) vanish and therefore
$\eta u^{a}$ is an RCV.

\section{Spacelike Ricci Collineation Vector}

Before we discuss the calculation some general results
can be presented for convenience on spacelike congruences that
will be used in this work. Let $\xi^{a}=\xi n^{a}$ where $n^a$ is
a unit spacelike vector ($n_a n^a=+1$) normal to the four velocity
vector $u^a$. The screen projection operator $P_{ab}=g_{ab}+u_a
u_b -n_a n_b$ projects normally to both $u^a$ and $n^a$ . Some
properties of this tensor are
\[P^{ab}u_{b}=P^{ab}n_{b}=0,\quad P^{a}_{c}P^{c}_{b}=P^{a}_{b},\quad
P_{ab}=P_{ba},\quad P^{a}_{a}=2.\] The $n_{a;b}$ can be decomposed
with respect to $u^a$ and $n^a$ as follows:
\begin{equation}\label{eq23}
n_{a;b}=A_{ab}+\stackrel{*}n_{a}n_{b}-\dot
n_{a}u_{b}+u_{a}\left[n^t u_{t;b}+(n^t\dot
u_{t})u_{b}-(n^t\stackrel{*}u_{t})n_{b}\right],
\end{equation}
where $\stackrel{*}s^{\; \ldots}_{\;\ldots}=s^{\;
\ldots}_{\;\ldots\; ;a} n^a$ and
$A_{ab}=P^{c}_{a}P^{d}_{b}n_{c;d}$. We decompose $A_{ab}$ into its
irreducible parts
\begin{equation}\label{eq24}
A_{ab}=S_{ab}+W_{ab}+\frac{1}{2}\stackrel{*}\theta P_{ab},
\end{equation}
where $S_{ab}=S_{ba},\; S^{a}_{a}=0$ is the traceless part of $A_{ab}$,
$\stackrel{*}\theta $ is the trace of $A_{ab}$, and $W_{ab}=-W_{ba}$ is the
rotation of $A_{ab}$. We have the relations:
\begin{eqnarray}\label{eq25}
S_{ab}&=&P^{c}_{a}P^{d}_{b}n_{(c;d)}-\frac{1}{2}\stackrel{*}\theta P_{ab},\\
W_{ab}&=&P^{c}_{a}P^{d}_{b}n_{[c;d]},\nonumber
\\\label{eq26}
\stackrel{*}\theta &=&P^{ab}n_{a;b}.
\end{eqnarray}
It is easy to show that in equation ~(\ref{eq23}) the $u^a$ term
in parenthesis can be written in a very useful form as follows:
\[-N_{a}+2\omega_{tb}n^t+P^{t}_{b}\dot n_{t},\]
where the vector

\begin{equation}\label{eq27}
N_{a}=P^{b}_{a}(\dot n_{b}-\stackrel{*}u_{b})
\end{equation}
is the Greenberg vector. Using ~(\ref{eq27}), equation
~(\ref{eq23}) becomes
\begin{equation}\label{eq28}
n_{a;b}=A_{ab}+\stackrel{*}n_{a}n_{b}-\dot
n_{a}u_{b}+P^{c}_{b}\dot n_{c}u_{a}+(2\omega_{tb}n^t-N_{b})u_{a}.
\end{equation}
The vector $N^a$ is of fundamental importance in the theory of
spacelike congruences. Geometrically the condition $N^a =0$ means
that the congruences $u^a$ and $n^a$ are two surface forming.
Kinematically it means that the field $n^a$ is "frozen in" along
the observers $u^a$.

Having mentioned a few basic facts on the spacelike congruences we
return to the computation of the Lie derivative of the Ricci
tensor using the field equations.

If Einstein's field equation ~(\ref{eq3}) are satisfied, then
string fluid with energy-momentum tensor ~(\ref{eq5}) admits an
RCV, $\xi^a =\xi n^a$ if and only if
\begin{eqnarray}
& &
q\omega_{at}n^t=\frac{1}{2}\rho_{s} N_{a},\label{eq29}\\
& & \rho_{s} S_{ab}=0,\label{eq30}\\ & &
q\left[\stackrel{*}n_{a}+(\ln\xi)_{,a}-(n_{t}\dot u^{t})n
_{a}\right]=0,\label{eq31}\\& & q\stackrel{*}\theta
=0,\label{eq32}\\& & \left(\xi q n^{a}\right)_{;a}=0.\label{eq33}
\end{eqnarray}

{\bf Proof:} From the definition of the Lie derivative it follows
that
\begin{equation}\label{eq34}
\pounds_{\xi n}R_{ab}=\xi\left[\stackrel{*}R_{ab}+2 n^c
R_{c(a}(\ln\xi)_{,b)}+2 R_{c(a} n^{c}_{;b)}\right]
\end{equation}
which, using Einstein's field equation~(\ref{eq3}) for string
fluid, may be rewritten as
\begin{eqnarray}\label{eq35}
\pounds_{\xi n}R_{ab} &= \xi\Bigg{\{} \stackrel{*}q (u_a u_b-n_a
n_b) +\stackrel{*}\rho_{s} P_{ab} -2
(\rho_{s}+q)\stackrel{*}n_{(a}n_{b)}
-2q n_{(a}(\ln\xi)_{,b)}\nonumber\\
&+2(\rho_{s}+q)\left[\stackrel{*}u_{(a}u_{b)}-n_{t}
u_{(a}u^{t}_{;b)}\right]+2\rho_{s} n_{(a;b)}\Bigg{\}}.
\end{eqnarray}
Suppose first that $\xi n^a$ is an RCV. Then  equation
~(\ref{eq1}) is satisfied. The right-hand side of equation
~(\ref{eq35}) is therefore zero and by contracting it in turn with
$u^a u^b$, $u^a n^{b}$, $u^a P^{bc}$, $n^a n^b$, $n^a P^{bc}$,
$P^{ab}$, and $P^{ac} P^{bd}-\frac{1}{2}P^{ab} P^{cd}$ the
following seven equations are derived:
\begin{eqnarray}
\stackrel{*}q+2q n_{a}\dot u^{a} =0,\label{eq36}\\
q\left[(\ln\xi)^{.}+\stackrel{*}n_{a}u^{a}\right]=0, \label{eq37}\\
\rho_{s}P^{b}_{a}\dot n_{b}-(\rho_{s}+q) P^{b}_{a}\stackrel{*}u_{b}+q
P^{b}_{a}n^{t}u_{t;b}=0, \label{eq38}\\
\stackrel{*}q +2q(\ln\xi)^{*}=0, \label{eq39}\\
q P^{b}_{a}\left[\stackrel{*}n_{b}+(\ln\xi)_{,b}\right]=0, \label{eq40}\\
\stackrel{*}\rho_{s}+\rho_{s}\stackrel{*}\theta =0, \label{eq41}\\
\rho_{s} S_{ab}=0.\label{eq42}
\end{eqnarray}

The energy momentum conservation equation will also be required.
For string fluid, the momentum conservation equation, which
follows from Einstein's field equations, is
\begin{equation}\label{eq43}
\stackrel{*}\rho_{s}=-(\rho_{s}+q)\stackrel{*}\theta.
\end{equation}
\begin{description}
\item [(i)] Condition ~(\ref{eq29}) is derived from ~(\ref{eq38}). We have
\begin{equation}\label{eq44}
n^t u_{t;b}=2
n^{t}u_{[t;b]}+\stackrel{*}u_{b}=-2\omega_{bt}n^t-(n_t \dot
u^{t})u_{b}+\stackrel{*}u_{b},
\end{equation}
and by substituting from ~(\ref{eq44}) into ~(\ref{eq38}),
~(\ref{eq29}) follows directly.

\item [(ii)] Condition ~(\ref{eq30}) is given by equation ~(\ref{eq42}).

\item [(iii)] To derive condition ~(\ref{eq31}) we first expand ~(\ref{eq40}) and
use ~(\ref{eq37}); this gives
\begin{equation}\label{eq45}
q\left[\stackrel{*}n_{a}+(\ln\xi)_{,a}-(\ln\xi)^{*}n_{a}\right]=0.
\end{equation}
If we subtract ~(\ref{eq39}) from ~(\ref{eq36}), then we have
\begin{equation}\label{eq46}
q(\ln\xi)^{*}=q n_{a}\dot u^{a}.
\end{equation}
If we substitute equation ~(\ref{eq46}) into equation
~(\ref{eq45}), then we have condition ~(\ref{eq31}).

\item [(iv)] To derive condition ~(\ref{eq32}), we substitute
equation ~(\ref{eq43}) into ~(\ref{eq41}), then we have condition
~(\ref{eq32}).

\item [(v)] Consider the final condition ~(\ref{eq33}).
From ~(\ref{eq26}), we have
\begin{equation}\label{eq47}
n_{a}\dot u^{a}=n^{a}_{;a}-\stackrel{*}\theta.
\end{equation}
Substitute ~(\ref{eq47}) into ~(\ref{eq46}); this gives
\begin{equation}\label{eq48}
q(\ln\xi)^{*}=q(n^{a}_{;a}-\stackrel{*}\theta).
\end{equation}
If one of the terms $q(\ln\xi)^{*}$ into equation ~(\ref{eq39}) is
replaced by ~(\ref{eq48}) and used condition ~(\ref{eq32}), then
~(\ref{eq39}) may be written as
\begin{equation}\label{eq49}
q_{,a}\xi n^{a}+q(\xi_{,a}n^{a}+\xi n^{a}_{;a})=0,
\end{equation}
from which ~(\ref{eq33}) follows directly.
\end{description}

Hence, if $\xi^a =\xi n^a$ is an RCV then conditions
~(\ref{eq29})-~(\ref{eq33}) are satisfied.

Conversely, suppose that ~(\ref{eq29})-~(\ref{eq33}) are satisfied
and Einstein's field equations hold. Using ~(\ref{eq28}) for
$n_{(a;b)}$, ~(\ref{eq30}) and ~(\ref{eq31}) for $q(\ln\xi)_{,a}$
equation ~(\ref{eq35}) becomes

\begin{eqnarray}\label{eq50}
\pounds_{\xi n}R_{ab}&=\xi\Bigg{\{}\stackrel{*}q u_a u_b
-(\stackrel{*}q +2q n_{t}\dot
u^{t})n_{a}n_{b}+(\stackrel{*}\rho_{s}+\rho_{s} \stackrel{*}\theta
)
P_{ab}\nonumber\\
&+2q\left[\stackrel{*}u_{(a}u_{b)}-n_{t}
u_{(a}u^{t}_{;b)}\right]-2\rho_{s} N_{(a}u_{b)}\Bigg{\}}.
\end{eqnarray}
Further, by using ~(\ref{eq44}) for $n^t u_{t;b}$ and
~(\ref{eq29}) for $q\omega_{at}n^t$ and by replacing
$\stackrel{*}\theta $ by $n_{t}\dot u^{t}$ with the aid of
~(\ref{eq26}), ~(\ref{eq50}) reduces to
\begin{equation}\label{eq51}
\pounds_{\xi n}R_{ab} = \xi\left\{\left(\stackrel{*}q+2q n_{t}\dot
u^{t}\right)(u_a u_b-n_{a}n_{b})
+(\stackrel{*}\rho_{s}+\rho_{s}\stackrel{*}\theta )P_{ab}\right\}.
\end{equation}
Now, $\stackrel{*}\rho_{s}$ is given equation ~(\ref{eq43}). To
obtain $\stackrel{*}q$ in terms of $n_{t}\dot u^{t}$  we use the
remaining condition ~(\ref{eq33}), which may be expanded as
\begin{equation}\label{eq52}
\stackrel{*}q+q(\ln\xi)^{*}+q n^{a}_{;a}=0.
\end{equation}
But if ~(\ref{eq31}) is contracted with $n^a$, equation
~(\ref{eq46}) is again derived. Therefore ~(\ref{eq52}) becomes
\begin{equation}\label{eq53}
\stackrel{*}q +2q n_{t}\dot u^{t}=0.
\end{equation}

It easily verified with the aid of ~(\ref{eq43}), ~(\ref{eq53}),
and condition ~(\ref{eq32}) that the right-hand side of
~(\ref{eq51}) vanishes and therefore $\xi^a =\xi n^a$ is an RCV.

\section{Results and Conclusions}

In the case of timelike Ricci collineation vectors
parallel to $u^a$, we have the following results:
\begin{description}
\item [(a)] In this case, it is easily verified that condition
~(\ref{eq9}) is the conservation law ~(\ref{eq4}) with
$\xi_{b}=\eta u_{b}$.

\item [(b)] Condition ~(\ref{eq8}) may be rewritten alternatively
either $q=0$ or $\dot u_{a}=(\ln\eta)_{,a}+\theta u_{a}$. If,
$q=0$ the energy-momentum tensor reduces
\[T_{ab}=\rho_{s}(u_{a}u_{b}-n_{a}n_{b})\]
which is pure string.
\end{description}
In the case of spacelike Ricci collineation vectors orthogonal to
$u^a$, we have the following results:

\begin{description}
\item [(a)] In this case, it is easily verified that ~(\ref{eq33})
is the conservation law ~(\ref{eq4}) for the string fluid with
$\xi_{b}=\xi n_b$.

\item [(b)] From equation ~(\ref{eq30}), we have
\begin{equation}\label{eq54}
\textrm{either}\quad \rho_{s}=0\quad \textrm{or} \quad  S_{ab}=0.
\end{equation}

\item [(c)] When $\omega=0$, equation ~(\ref{eq29}) reduces to
\begin{equation}\label{eq55}
\rho_{s} N_{a}=0,
\end{equation}
and hence either $\rho_{s}=0$ or $N_{a}=0$. When $N_{a}=0$, the integral curves
$n^a$ are material curves and string fluid form two surface. When $\rho_{s}=0$,
strings disappear.

\item [(d)] When $\omega\neq 0$, equation ~(\ref{eq29}) reduces to
\begin{equation}\label{eq56}
q\omega_{at}n^t=\frac{1}{2}\rho_{s} N_{a}.
\end{equation}
\end{description}
\begin{description}
\item [(i)] If $N_{a}=0$, then equation ~(\ref{eq56}) reduces to
\begin{equation}\label{eq57}
q\omega_{at}n^t=0
\end{equation}
and hence if $q\neq 0$ then $\omega_{at} n^{t}=0$ and since
$\omega_{at}=\eta_{atrs}\omega^{r}u^{s}$ we find by contracting
~(\ref{eq57}) with $\eta^{abcd}\omega_{c}u_{d}$ that
\begin{equation}\label{eq58}
n^{a}=\left[(\omega_{t}n^{t})/\omega^{2}\right]\omega^{a}.
\end{equation}
Since both $n^{a}\neq 0$ and $\omega^{a}\neq 0$ it follows that
$n^{a}=\pm\omega^{a}/\omega$. \item [(ii)] If the integral curves
of $n^a$ are material curves in the fluid then $N^{a}=0$. Hence,
since $q\neq 0$, from ~(\ref{eq46}) $\omega_{at} n^{t}=0$ and
$n^{a}=\pm\omega^{a}/\omega$. \item [(iii)] If
$n^{a}=\pm\omega^{a}/\omega$ and $q\neq 0$ then from equation
~(\ref{eq56}), $N^{a}=0$ and the integral curves of $n^a$ are
material curves.
\end{description}

\end{document}